# Towards Reliable WMH Segmentation under Domain Shift: An Application Study using Maximum Entropy Regularization to Improve Uncertainty Estimation

Franco Matzkin* [a], Agostina Larrazabal[c], Diego H Milone[a], Jose Dolz[b], Enzo Ferrante* [d]

[a] *Institute for Signals, Systems and Computational Intelligence, sinc(i) CONICET-UNL, Santa Fe, Argentina.*

[b] *Laboratory for Imagery, Vision and Artificial Intelligence, LIVIA, ETS, Montreal, Canada.*

[c] *Tryolabs, Uruguay*

[d] *Institute of Computer Sciences, ICC, CONICET-Universidad de Buenos Aires, Ciudad Autónoma de Buenos Aires, Argentina*

\* Corresponding authors:

- Franco Matzkin. E-mail: fmatzkin@sinc.unl.edu.ar. Address: Ciudad Universitaria UNL, Ruta Nacional Nº 168, km 472.4, FICH, 4to Piso. Santa Fe, Argentina. Phone: +54 (0342) 4575233 (ext: 195)

- Enzo Ferrante. E-mail: eferrante@dc.uba.ar. Address: Pabellón Cero+Infinito – Ciudad Universitaria. C1428EGA C.A.B.A. Tel./Fax (54.11) 5285-9732

## Abstract

**Background**

Accurate segmentation of white matter hyperintensities (WMH) is crucial for clinical decision-making, particularly in the context of multiple sclerosis. However, domain shifts, such as variations in MRI machine types or acquisition parameters, pose significant challenges to model calibration and uncertainty estimation. This comparative study investigates the impact of domain shift on WMH segmentation, proposing maximum-entropy regularization techniques to enhance model calibration and uncertainty estimation. The purpose is to identify errors appearing after model deployment in clinical scenarios using predictive uncertainty as a proxy measure, since it does not require ground-truth labels to be computed.

**Methods**

We conducted experiments using a classic U-Net architecture and evaluated maximum entropy regularization schemes to improve model calibration under domain shift on two publicly available datasets: the WMH Segmentation Challenge and the 3D-MR-MS dataset. Performance is assessed with Dice coefficient, Hausdorff distance, expected calibration error, and entropy-based uncertainty estimates.

**Results**

Entropy-based uncertainty estimates can anticipate segmentation errors, both in-distribution and out-of-distribution, with maximum-entropy regularization further strengthening the correlation between uncertainty and segmentation performance, while also improving model calibration under domain shift.

**Conclusions**

Maximum-entropy regularization improves uncertainty estimation for WMH segmentation under domain shift. By strengthening the relationship between predictive uncertainty and segmentation errors, these methods allow models to better flag unreliable predictions without requiring ground-truth annotations. Additionally, maximum-entropy regularization contributes to better model calibration, supporting more reliable and safer deployment of deep learning models in multi-center and heterogeneous clinical environments.

## Highlights

- Entropy-based uncertainty estimates can be used as a proxy for segmentation errors.
- Maximum-entropy regularization improves model calibration and uncertainty quantification under domain shift in WMH segmentation.
- Models trained with maximum-entropy regularization achieve stronger alignment between uncertainty and segmentation errors.
- Validation performed on multicenter WMH datasets highlights robustness to different imaging conditions.

## 1. Introduction

Accurate segmentation in medical imaging is crucial for a variety of clinical applications, from computer-aided diagnostics to treatment planning (Yang and Yu, 2021). In the context of Multiple Sclerosis (MS) research, the segmentation of hyperintensity areas, identifiable on head MRI scans, are indicative of pathological changes in the brain and are closely associated with MS pathology. Developing robust automated segmentation methods is crucial to improve the understanding of white matter hyperintensities (WMH) and enhancing diagnosis, monitoring, and treatment strategies for MS patients, making WMH segmentation a key predictor (Palladino et al., 2020). Accurate and reliable WMH segmentation directly impacts patient care and clinical decision-making, as it helps in estimating lesion load, an important marker for disease progression and treatment response (Chaves et al., 2024).

This task is usually approached using deep learning strategies based on convolutional neural networks (CNNs) (Tran et al., 2022). Models based on CNNs, known for their outstanding performance in segmentation tasks, heavily rely on consistent distributions between training and test datasets. When confronted with changes in distribution, such as variations in MRI machine types or acquisition parameters across different medical centers, a phenomenon known as domain shift occurs, usually leading to a decline in segmentation accuracy. This presents a significant challenge as it compromises the model's ability to generalize effectively across diverse imaging scenarios. In addition to compromising the discriminative performance of the model, domain shift can also impact its calibration (Ricci Lara et al., 2023; Mosquera et al., 2024; Ovadia et al., 2019). Calibration, which refers to the alignment between predicted probabilities and observed outcomes, is essential for accurate decision-making (Sambyal et al., 2023). When faced with domain shift, the model predictions could become less calibrated in the target domain, potentially misleading the clinician's interpretation of the results. Poor calibration can lead to overconfidence in wrong decisions or unnecessary doubts about correct ones. While one would expect the probabilistic outputs of CNN segmentation models to be affected by domain shift, manifesting higher uncertainty in the predictions, this is not usually the case. Instead, models tend to remain overconfident even in situations where predictions are wrong (e.g. producing predictions close to 0 -background- or 1 -lesion- in a binary lesion segmentation scenario, instead of assigning values close to 0.5 which would better reflect uncertainty about the unknown data distribution).

Therefore, addressing domain shift is not only important to ensure accurate segmentation, but also plays a vital role in maintaining the calibration of the model across domains, ultimately enhancing its utility in clinical practice. In this work, we are interested in quantifying model uncertainty under domain shift scenarios, a concept closely related to model calibration. In cases when we go from in-distribution (ID) data samples, which are similar to the training data, to out-of-distribution (OOD) samples, which deviate from the training data distribution, uncertainty quantification (UQ) can allow us to flag segmentation cases which require intervention (Mehrtash et al., 2020).

In the context of medical imaging, models trained with classical loss functions (such as the popular Cross Entropy -CE- or soft Dice loss (Milletari et al., 2016)) may exhibit overconfidence in their predictions when faced with OOD data, leading to suboptimal outcomes. An example is shown in Figure 1 (central column), where a WMH segmentation prediction generated by a model trained using a classical pixel level CE as the loss function, shows the label likelihood close to one across the entire segmented area. However, it would be more beneficial for the model to express uncertainty in areas where less consensus between raters could be expected, such as at lesion boundaries or in small, isolated lesions distant from larger lesion areas (as in the right column). This discrepancy underscores the need for more sophisticated loss functions and training strategies that can effectively address domain shift challenges in medical imaging applications, encouraging the model to doubt in OOD scenarios, instead of producing overconfident predictions.

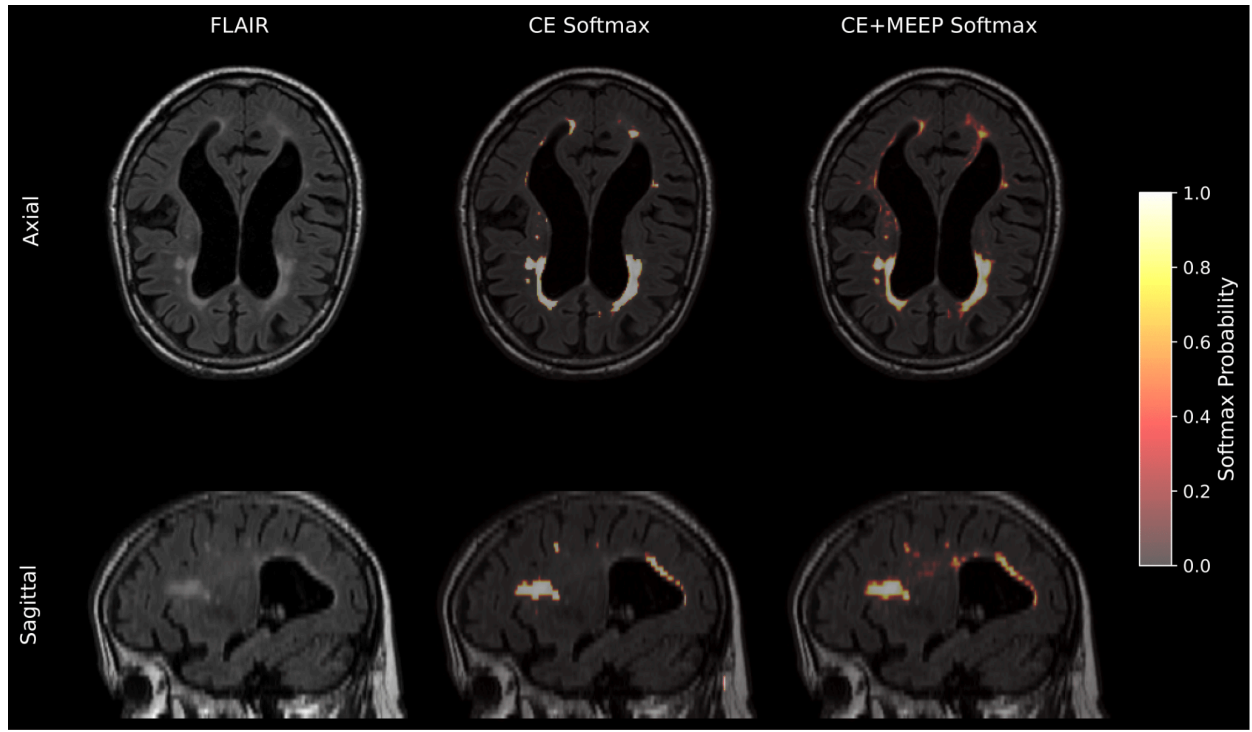

Figure 1: Comparison of White Matter Hyperintensity (WMH) segmentation on axial and sagittal FLAIR MRI from a multiple sclerosis patient. The input FLAIR image (left) and an overconfident output from a CE Softmax model (center) are contrasted with the result from CE+MEEP Softmax (right). This latter approach yields more detailed probabilistic segmentations, capturing uncertainty more effectively through intermediate values—especially around lesion boundaries and in small WMH regions.

Previous work has proposed the use of regularization methods to discourage overconfident predictions. In the context of classification problems, Pereyra et al. (2017) proposed to increase entropy in the probabilistic output (i.e. preventing peaked distributions and promoting uniformity) of classification models by incorporating an additional regularization term to the loss function, representing the negative entropy of the output probability. Since confident predictions correspond to output distributions that have low entropy, this regularization term that prevents peaked distributions was shown to help avoid overconfidence for ID data. However, it was not evaluated under distribution shift scenarios. This idea was further refined in (Larrazabal et al., 2023) where, instead of penalizing low entropy for all predictions, only the erroneous ones were penalized, resulting in more accurate segmentations for ID data.

So far, the use of maximum entropy methods for image segmentation has mostly been limited to ID data. At the same time, previous work (Nair et al., 2020) investigated the use of entropy as a measure for uncertainty quantification in the context of WMH segmentation. However, they did not explore maximum entropy methods to enhance these estimates, nor did they address the implications of distribution shifts, which are a critical issue in multi-centric scenarios. Here we study maximum entropy methods to improve UQ *under distribution shifts* in WMH segmentation. In particular, we will examine whether these models can maintain accurate uncertainty estimation when confronted with changes in data distribution, crucial for reliable decision-making in clinical settings. Additionally, we will explore model calibration under OOD scenarios, providing insights into the effectiveness of the maximum entropy methods in detecting erroneous cases. By assessing the model performance across various medical centers and imaging scenarios, our goal is to uncover its adaptation and generalization capacity in diverse clinical environments, ultimately aiming to provide valuable guidance for integrating deep learning models into clinical practice and advancing patient care outcomes.

**Contributions:** Our main contributions are threefold: 1) we investigate the impact of domain shift on model calibration for WMH segmentation, 2) we propose the use of maximum entropy regularization for improving uncertainty estimates in WMH segmentation under domain shift, and 3) we assess the correlation between uncertainty and segmentation errors in this scenario. By achieving these goals, we aim to enhance the reliability and clinical applicability of deep learning models in the context of WMH segmentation for MS patients. Specifically, we hypothesize that higher entropy values will correlate with lower Dice scores, particularly under domain shift conditions, enabling entropy-based uncertainty estimates to serve as reliable proxies for segmentation performance. To validate this hypothesis, we systematically evaluate existing entropy-based regularization methods on multicentric MRI datasets acquired under varying scanning protocols and patient populations. In our experiments, maximum entropy regularization methods indeed improved uncertainty estimation and calibration under domain shift.

## 2. Materials and methods

Let us say we have a segmentation model $S: X \rightarrow Y$ that, given an image $X$, returns a probabilistic voxel level segmentation map $Y$, as $Y = S(X)$. For every voxel $i$, $Y$ will assign a probability $y_i$ for the WMH lesion class, and $1 - y_i$ will be the probability of healthy tissue. Without loss of generality, in our case the model $S$ is an encoder-decoder convolutional neural network which follows a U-Net architecture (Ronneberger et al., 2015). Note that this formulation is model-agnostic, and hence other architectures could also be considered. Given the probabilistic segmentation map, we aim to estimate voxel-level uncertainty. In this study, we focus on predictive entropy as the uncertainty metric.

### 2.1 Entropy-based uncertainty estimation

Various methods have been proposed for estimating uncertainty in medical image segmentation, including Monte Carlo Dropout (Gal and Ghahramani, 2016), model ensembling, and Probabilistic U-Net (Kohl et al., 2018). In this work, we focus on predictive entropy, a widely adopted approach (Czolbe et al., 2021; Nair et al., 2020) due to its simplicity and interpretability.

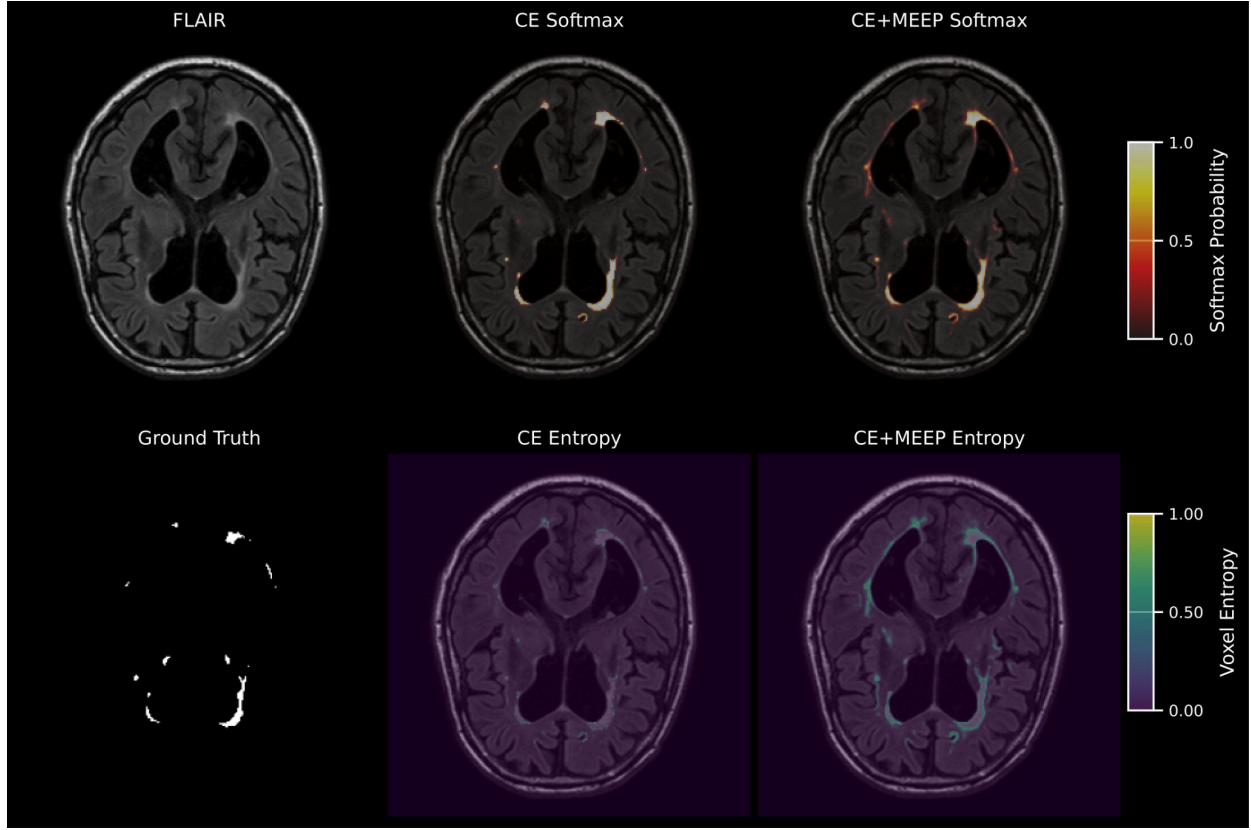

Figure 2: Input FLAIR MRI (top left) and ground truth segmentation (bottom left) for White Matter Hyperintensities (WMH). These are shown alongside softmax probability outputs from CE Softmax (top center) and CE+MEEP Softmax (top right) models, and their respective voxel entropy maps: CE Entropy (bottom center) and CE+MEEP Entropy (bottom right). Notably, the CE+MEEP Entropy maps more distinctly highlights uncertainty in small WMH lesions visible in the ground truth, compared to the CE Entropy map.

Uncertainty in model predictions can be estimated using predictive entropy. For binary segmentation, the binary entropy of the segmented region has been employed to provide insights into the confidence levels associated with the predictions (Czolbe et al., 2021; Mehrtash et al., 2020; Nair et al., 2020). Given a Bernoulli probability distribution parameterized by $p$, its binary entropy is defined as

$$H_b(p) = -p \, log_2(p) - (1-p)log_2(1-p),$$

where $p$ stands for the probability that a voxel or data point belongs to the foreground class, which, in case of WMH segmentation, is the probability associated with the lesion class. The binary entropy $H_b$ could range from 0 to 1: when $H_b = 0$, the outcome is entirely predictable, and when $H_b = 1$ it is completely unpredictable or random. In a binary segmentation scenario, if a model assigns a probability close to 1 for a voxel belonging to the target class, then the entropy will be very low, indicating high confidence. Alternatively, if the model assigns a probability of 0.5, the entropy will be maximum, indicating high uncertainty (see Figure 2). This allows practitioners to identify uncertain regions, potentially requiring further inspection or intervention, thus enhancing the model reliability and interpretability.

## 2.2 Improving entropy-based uncertainty estimation via maximum entropy methods

We propose three strategies to promote higher entropy distributions and evaluate their effectiveness in terms of uncertainty estimation under domain shift. Such strategies are implemented as an additional term in the loss function for training the neural network. In general, we will train our models using the following loss function:

$$L = L_{seg}(Y, \widehat{Y}) + \lambda L_{reg}(Y),$$

where $L_{seg}$ is the data term (either cross entropy or soft Dice loss) computed by comparing the predicted segmentation mask $Y$ with the ground-truth label $\widehat{Y}$, and $L_{reg}$ is a regularization term defined to encourage high entropy. In what follows, we introduce three alternatives for this regularization term.

### 2.2.1 Overall confidence penalty

As previously discussed, overconfident models tend to assign all probability into a single class. To avoid such behavior, we first propose to encourage high entropy for all voxel predictions. We follow the idea introduced by (Pereyra et al., 2017) in the context of image classification, adapting it to the context of image segmentation. Thus the entropy of *all* voxel predictions $y_i \in Y$ in the predicted segmentation mask $Y$ is computed, defining the regularization term as

$$L_a(Y) = - H_b(Y) = - \sum_{y_i \in Y} - y_i \log_2(y_i) - (1 - y_i)\log_2(1 - y_i).$$

This term is added in the the overall loss function, encouraging maximum entropy for all voxel predictions: $L(Y, \widehat{Y}) = L_{seg}(Y, \widehat{Y}) + \lambda L_a(Y)$. This approach systematically enforces higher entropy in the outputs of the model, acting as a strong regularizer and improving generalization by reducing overconfidence even in correct predictions.

### 2.2.2 Maximum entropy on erroneous predictions

The term defined in the previous section penalizes high confidence for all voxel predictions. However, if a prediction is correct, in principle there is nothing wrong with the model being confident about it. Indeed, we argue that one would like to avoid overconfident predictions especially in cases where those predictions are wrong. Thus, we resort to the maximum entropy on erroneous predictions (MEEP) regularizer, $L_m(Y_w)$, which penalizes low entropy only for erroneous predictions. We will use $\mathbf{y}_w$ to define the set of voxels whose label was incorrectly predicted, and hence we can define the regularizer as

$$L_m(Y_w) = - H_b(Y_w) = - \sum_{y_i \in Y_w} - y_i \log_2(y_i) - (1 - y_i)\log_2(1 - y_i).$$

This regularizer will penalize low entropy (i.e. peaky) distributions only when the predictions are wrong, which intuitively encourages uniform predictions in highly uncertain situations. In particular, we hypothesize that this term will help in domain shift scenarios due to changes in intensity distributions when facing multicentric datasets. Similarly as before, we will add this term to the overall loss function, encouraging maximum entropy only for voxels which were wrongly predicted, resulting in the following loss $L(Y, \hat{Y}) = L_{seg}(Y, \hat{Y}) + \lambda L_m(Y_w)$.

### 2.2.3 Maximum entropy on erroneous predictions via KL divergence

We evaluate a third approach where we also encourage high entropy in erroneous predictions but following a different strategy. Instead of subtracting the entropy of misclassified voxels from the overall loss function, we introduce a regularization term to encourage their predictions to be uniformly distributed, by minimizing the Kullback-Leibler (KL) divergence with respect to a uniform distribution. The KL divergence $D_{KL}(Q||P)$ provides a notion of difference between two probability distributions $P$ and $Q$. Since the uniform distribution has maximum entropy, we will minimize the difference between the predicted distribution for misclassified voxels $Y_w$ and the uniform distribution Q, by adding a regularization term $L_{KL}(Y_w) = - D_{KL}(Q||Y_w)$, resulting in the loss function: $L(Y, \hat{Y}) = L_{seg}(Y, \hat{Y}) + \lambda L_{KL}(Y_w)$. Note that although $L_{KL}(Y_w)$ and $L_m(Y_w)$ drive $Y_w$ towards a uniform distribution, their gradient dynamics differ, resulting in different effects on the neural weight updates during training. In this study, we conduct an experimental analysis to determine which term yields better UQ under domain shift.

## 2.3 Metrics and evaluation protocols

Here we are interested in assessing how WMH segmentation models behave under domain shift, improve their performance both in terms of discrimination and calibration, and understand if the entropy of the predictions can be used as a proxy to anticipate potential failures. These aspects provide a comprehensive understanding of the overall performance and its suitability for real-world applications. In what follows, we describe the metrics that are used to evaluate each of these aspects.

### 2.3.1 Discrimination metrics

Discriminative ability is achieved when the model can effectively distinguish between different classes. For the segmentation tasks, the **Dice Coefficient** was used. This widely used metric measures the overlap between the predicted segmentation and the ground truth. It is calculated as

$$Dice = \frac{2|G \cap P|}{|G|+|P|},$$

where $|G \cap P|$ represents the number of elements common to both the ground truth set $G$ and the predicted set $P$, and $|\cdot|$ denote the number of elements in the set.

### 2.3.2 Calibration metrics

Calibration metrics are crucial for assessing how well the predicted probabilities of a model align with actual outcomes. Previous studies have shown that segmentation models trained with Dice loss tend to be overconfident (Yeung et al., 2023; Murugesan et al., 2023), while cross-entropy training typically leads to better calibrated models (Mehrtash et al., 2020).

Among calibration metrics, the Expected Calibration Error (ECE) is useful for assessing the reliability of probability estimates. To calculate it, we first allocate each voxel prediction to a bin, depending on the predicted probability value. Here we consider bin separation of $0.1$, resulting in $M = 10$ bins of the form $\{B_o = [0, 0.1), B_1 = [0.1, 0.2) \dots B_{10} = [0.9, 1]\}$. ECE is then calculated as

$$ECE = \sum_{m=1}^{M} \frac{|B_m|}{n} |acc(B_m) - conf(B_m)|,$$

where $|B_m|$ is the number of samples in bin $B_m$, $n$ is the total number of samples, $acc(B_m)$ is the accuracy of voxels in bin $B_m$, and $conf(B_m)$ is the average confidence (predicted probability value) of bin $B_m$. This metric captures the average discrepancy between predicted probability and actual accuracy across all bins.

Another essential tool for assessing calibration is the **reliability plot**. This graphical representation plots the average predicted probability, $p$, against the actual fraction of positives, $f_p$, for each bin. Ideally, the points in a reliability plot should lie on the line $p = f_p$, indicating perfect calibration where the predicted probability matches the observed frequency of the event. This visualization helps identify areas where the model is overconfident or underconfident in its predictions.

By incorporating these metrics, we can comprehensively evaluate both the discriminative power and the calibration quality of machine learning models, ensuring their reliability and effectiveness in clinical practice.

### 2.3.3 Uncertainty quantification protocols

To evaluate the relationship between segmentation performance and uncertainty estimates, we computed the Pearson correlation between the average foreground entropy and the Dice coefficient across scans. For each case, we first filtered voxels classified by the model as foreground (predicted probability > 0.5) and then computed the mean entropy over these voxels. This approach simulates a clinical scenario where ground-truth labels are unavailable, focusing the uncertainty analysis on the model's positive predictions.

### 2.3.4 Evaluation on different lesion sizes

Previous work has shown that WMH segmentation methods tend to present lower quality for smaller lesions (Chaves et al., 2024). Thus, one would expect that entropy-based uncertainty estimates present higher values for patients with smaller lesion load. We thus examine in Section 3.2 how uncertainty varies with lesion size in a comparative analysis, grouping the lesions according to their volume (smaller than 5 mL, between 5 mL and 15 mL and bigger than 15 mL).

## 2.4 Datasets

This retrospective study analyzed two WMH segmentation datasets:

**White Matter Hyperintensity (WMH) Segmentation Challenge:** WMH Segmentation Challenge dataset consists of brain MR images (T1 and FLAIR) with manual annotations of WMH. The dataset includes 60 training sets of T1/FLAIR images from three different institutions, annotated by experts in WMH scoring and 110 test sets from five different scanners. The dataset was derived from patients with various degrees of aging-related degenerative and vascular pathologies to ensure generalizability of segmentation methods across scanners and patient variability. The participants had a mean age of approximately 70 years (70.1 ± 9.3 years), with an equal gender distribution (50% male). WMH burden varied widely, with mean WMH volume of 16.9 ± 21.6 ml and a mean lesion count of 62 ± 35 lesions per subject. This dataset was created as part of the WMH Segmentation Challenge, associated with MICCAI 2017, and was active from 2017 to 2022. The challenge aimed to evaluate and compare methods for the automatic segmentation of WMH of presumed vascular origin. Participants trained their models on the provided training data and submitted their methods for evaluation using the unreleased test data. Results of this challenge have been published in (Kuijf et al., 2019).

**3D MR Image Database of Multiple Sclerosis Patients with White Matter Lesion Segmentations (3D-MR-MS):** The 3D-MR-MS dataset (Lesjak et al., 2018) comprises magnetic resonance (MR) images from 30 patients with multiple sclerosis (MS), acquired at the University Medical Center Ljubljana. The dataset includes co-registered and bias-corrected T1-weighted (T1W), contrast-enhanced T1-weighted (T1WKS), T2-weighted (T2W), and FLAIR images, as well as corresponding brain masks and intra-study transform parameters. The patients had a median age of 39 years (range: 25 to 64), with a female-to-male ratio of 23:7. The dataset is designed to support research in automated lesion segmentation for neurodegenerative diseases like MS. Lesion burden varied significantly, with a total of 3316 lesions segmented and an overall lesion volume (total lesion load, TLL) of 567 ml. The median lesion volume per subject was 15.2 ml (range: 0.337–57.5 ml, interquartile range: 31.1 ml). Lesion sizes ranged from 2 µl to 250 µl (5th to 95th percentile).

Our analysis utilized existing MRI scans and their corresponding manual WMH segmentations to develop and evaluate the proposed methods for uncertainty estimation in WMH segmentation. To ensure consistency, we applied identical preprocessing steps to both datasets, including resampling images to match their spatial resolutions, z-score intensity standardization, and N4 bias field correction (already provided for the 3D-MR-MS dataset).

## 2.5 WMH segmentation model details

For all experiments in this study, we employed a 3D U-Net architecture (Ronneberger et al., 2015) for WMH segmentation, implemented using the MONAI framework (Cardoso et al., 2022). The model was designed for 3D volumetric MRI data, accepting two input channels (FLAIR and T1-weighted images) and producing two output channels representing the background and WMH classes. The network consisted of four downsampling/upsampling levels, with feature channels set to (8, 16, 32, 64). Downsampling was achieved using 2×2×2 strided convolutions.

A dropout rate of 0.2 was applied within the network during training for regularization. The model was optimized with the Adam method, using an initial learning rate of 0.001 and no weight decay. Training was patch-based, with a batch size of 64 patches of size 32×32×32 voxels extracted from the input volumes, and proceeded for up to 800 epochs.

Regularization weights for the different regularization terms were selected individually for each strategy through grid search, balancing segmentation performance and the quality of uncertainty estimation.

Inference was also performed using a patch-based sliding window approach with the same patch size (32×32×32), aggregating predictions to reconstruct full-volume segmentations. This standardized model and preprocessing configuration (described in Section 2.4) provided a robust baseline, allowing for the evaluation of regularization strategies on model performance, calibration, and uncertainty estimation under domain shift.

## 3. Results

In this section, we empirically evaluate the proposed methods, investigating the relationship between model confidence, segmentation quality, and robustness of the model when exposed to OOD samples. We use the previously discussed White Matter Hyperintensity (WMH) Segmentation Challenge dataset (which is considered to be ID) and the 3D MR Image Database of Multiple Sclerosis Patients (3D-MR-MS), considered to be OOD.

### 3.1 Entropy as a proxy for error prediction in domain shift scenarios

We evaluated the relationship between segmentation performance and uncertainty estimates by analyzing the Pearson correlation between average foreground entropy (as described in Section 2.3.3) and Dice scores across scans. Figure 3 presents scatter plots of entropy as a function of Dice for both ID and OOD data across the four strategies: cross-entropy (CE), CE regularized with Maximum Entropy on Erroneous Predictions (CE$_{MEEP}$), CE regularized with Kullback-Leibler divergence (CE$_{KL}$), and CE with Maximum Entropy on All Predictions (CE$_{MEALL}$). Linear regression lines are fitted to each set of data points, revealing distinct trends for each loss function, regardless of the medical center. The Pearson correlation coefficient is provided for each loss function.

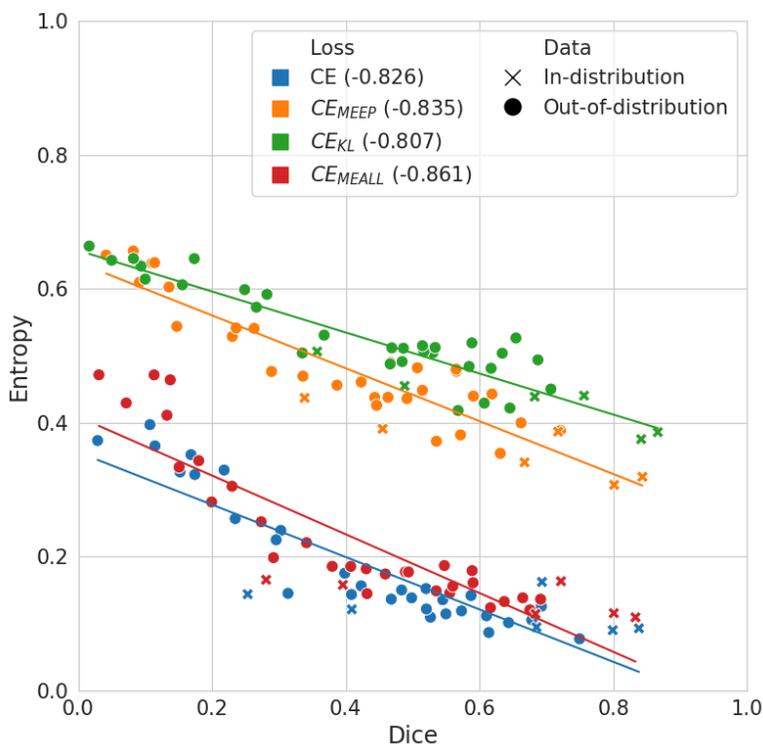

Figure 3: Scatter plot comparing entropy of foreground predictions and Dice coefficient, per image, for ID and OOD patients. Pearson correlation coefficient between entropy and Dice is shown in parenthesis in the legend box. It can be observed that entropy estimates for MEEP and KL yield better anti-correlation, thus serving as predictors of potential failures.

Although the differences in Pearson correlation coefficients are not very large, consistent trends are observed: regularization methods targeting uncertainty improvement (CE$_{MEEP}$ and CE$_{KL}$) systematically achieve stronger negative correlations between entropy and Dice scores compared to standard cross-entropy (CE). This indicates that entropy-based regularization leads to uncertainty estimates that more reliably reflect segmentation performance, supporting their use as practical predictors of potential failures, particularly under domain shift.

Across all loss functions, a negative correlation is observed between Dice and entropy, indicating that higher segmentation quality is generally associated with lower uncertainty. However, the strength of this correlation varies across loss functions, with $CE_{MEEP}$ and $CE_{KL}$ exhibiting stronger negative correlations (−0.835 and −0.807, respectively) compared to CE (−0.826) and $CE_{MEALL}$ (−0.861). Specifically, the Pearson correlation coefficients between average foreground entropy and Dice were −0.826 for CE, −0.835 for $CE_{MEEP}$, −0.807 for $CE_{KL}$, and −0.861 for $CE_{MEALL}$. This suggests that $CE_{MEEP}$ and $CE_{KL}$ may provide more reliable uncertainty estimates, as their entropy values more closely track the actual segmentation performance.

To further investigate the behavior of uncertainty estimates under domain shift, we examine their distribution across different types of prediction errors. **Figure 4** presents a scatterplot where each point represents a voxel, color-coded based on the classification outcome: true positives (TP), true negatives (TN), false positives (FP), and false negatives (FN). In addition, out-of-distribution (OOD) data are indicated by blue bars, while in-distribution (ID) data are shown with orange bars.

In the case of TP, CE and $CE_{MEALL}$ exhibit the lowest uncertainty, while $CE_{MEEP}$ and $CE_{KL}$ yield higher uncertainty, both for ID (blue) and OOD (orange) cases. For TN, a similar behavior is observed, although less dispersed, with uncertainty medians close to zero for all methods. As expected, FP exhibit higher uncertainties since the model is making incorrect predictions. Notably, $CE_{MEEP}$ and $CE_{KL}$ offer the highest uncertainty for these cases, both in and out of distribution. This heightened uncertainty for FP is desirable, as it allows for the identification of potentially erroneous segmentations, particularly in the challenging OOD setting where the model is more likely to find unfamiliar data distributions. FP often occur in regions with ambiguous image characteristics, making it difficult for the model to confidently distinguish them from TP. Finally, for FN, $CE_{MEEP}$ and $CE_{KL}$ again show higher uncertainty, indicating their ability to express doubt when the model is incorrect.

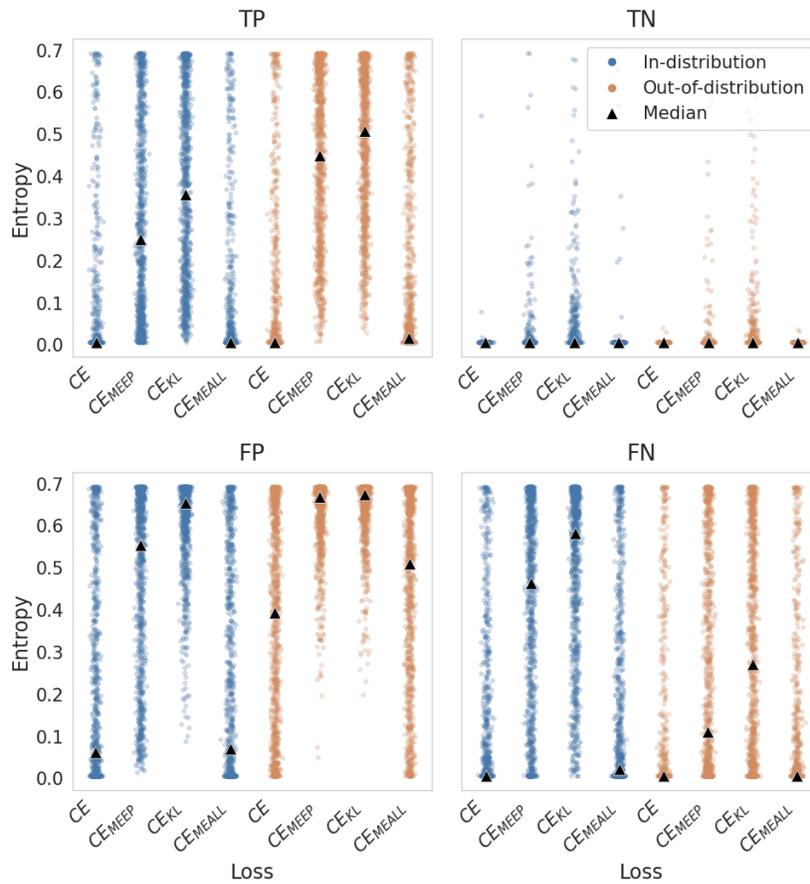

Figure 4: Distribution of uncertainty estimates across different prediction outcomes (True Positives, True Negatives, False Positives, False Negatives) for various training strategies under ID and OOD scenarios. Each point represents a voxel, with blue indicating ID data and orange representing OOD data. The x-axis shows different training strategies, while the y-axis represents entropy values. Black triangles denote median entropy values. This visualization allows for comparison of uncertainty behaviors across different loss functions, revealing how methods like $CE_{MEEP}$ and $CE_{KL}$ tend to yield higher uncertainties, particularly for false positives and false negatives, in both ID and OOD settings.

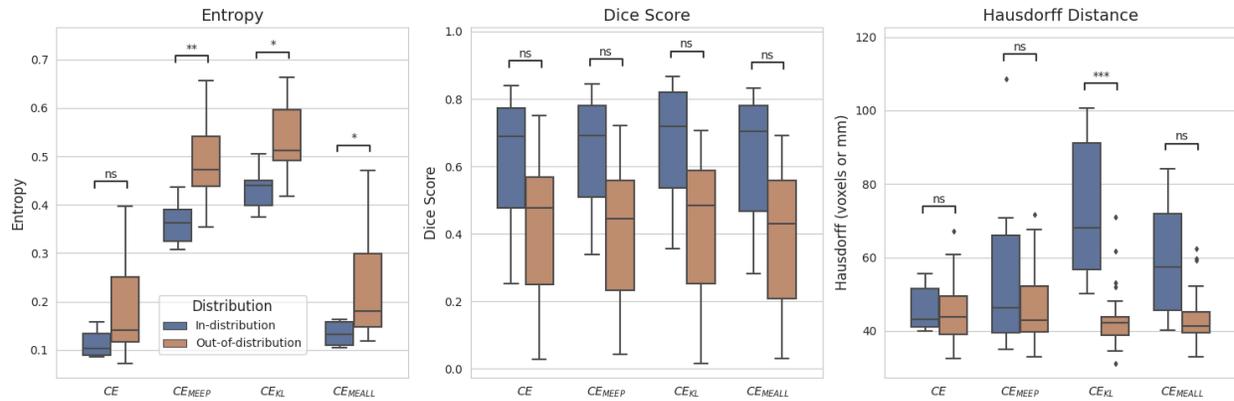

Figure 5: Boxplots comparing metrics across in-distribution (ID) and out-of-distribution (OOD) data for different loss functions. (Left): Average entropy for voxels predicted as positive, showing a general increase in uncertainty under domain shift, especially for $CE_{MEEP}$ and $CE_{KL}$. (Middle): Dice score performance across loss functions, with ID scores consistently higher than OOD scores. (Right): Hausdorff distances illustrating boundary localization performance across ID and OOD cases. Statistical significance is indicated where applicable according to the Mann–Whitney U test.

To gain deeper insights into how maximum entropy regularizers affect the uncertainty estimates, we first analyze entropy levels across ID and OOD data, as shown in Figure 5. As stated previously, the outcomes display two distinctive patterns: standard cross-entropy (CE) and $CE_{MEALL}$ exhibit lower entropy values, (i.e. which translate into higher confidence in their predictions). Conversely, $CE_{MEEP}$ and $CE_{KL}$ demonstrate elevated entropy levels, particularly for OOD data, suggesting increased sensitivity to domain shift and a greater ability to capture uncertainty in challenging scenarios.

A Mann-Whitney U test confirms this observation, revealing statistically significant differences in entropy levels between ID and OOD samples for $CE_{MEEP}$ and $CE_{KL}$, further supporting their effectiveness in distinguishing between the two scenarios. This ability to differentiate between ID and OOD data based on uncertainty estimates is crucial for identifying unreliable predictions and ensuring the model's robustness in real-world clinical settings.

## 3.2 Uncertainty and lesion size analysis

**Figure 6** shows that smaller lesions tend to have higher entropy across all loss functions. This observation aligns with the difficulty of reaching expert consensus on ground-truth labels for smaller lesions, as their subtle appearance can make them difficult to identify and delineate. Larger lesions are generally associated with lower entropy values, indicating higher model confidence, and this tendency is consistently observed for both ID and OOD cases.

Quantitatively, with CE the median entropy for small lesions (<5 mL) was approximately 0.58, compared to 0.23 for large lesions (>15 mL), illustrating the decrease in model uncertainty with increasing lesion size. Notably, the $CE_{MEEP}$ regularization strategy specifically targets these smaller lesions by pushing uncertainty levels toward the maximum, reflecting the inherent ambiguity and potential for disagreement in these cases. This targeted approach could be particularly valuable in clinical practice, as it allows the model to flag its own limitations and prompt further investigation or consultation for uncertain, small lesions.

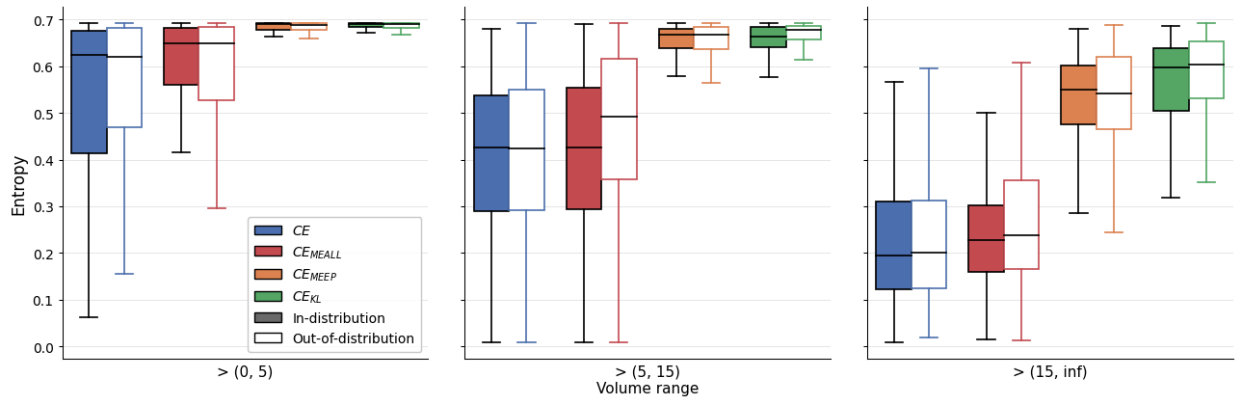

Figure 6: Boxplots comparing average entropy for voxels predicted as positive across different strategies in three lesion volume ranges. The plot distinguishes between ID (filled boxes) and OOD (unfilled boxes) data. We observe that larger lesion volumes are generally associated with lower entropy, confirming that it can serve as an indicator of model uncertainty. Notably, this tendency is conserved for both ID and OOD cases.

## 3.3 Model calibration in domain shift scenarios

Finally, to assess the impact of domain shift on model calibration, we analyze reliability diagrams and ECE for each loss function considering both ID and OOD scenarios (**Figure 7**).

In the ID scenario, $CE_{MEEP}$ outperforms other losses in terms of Expected Calibration Error (ECE), while in the OOD scenario, all loss functions exhibit poorer calibration, except for the KL-based loss, which demonstrates superior calibration and robustness to domain shift.

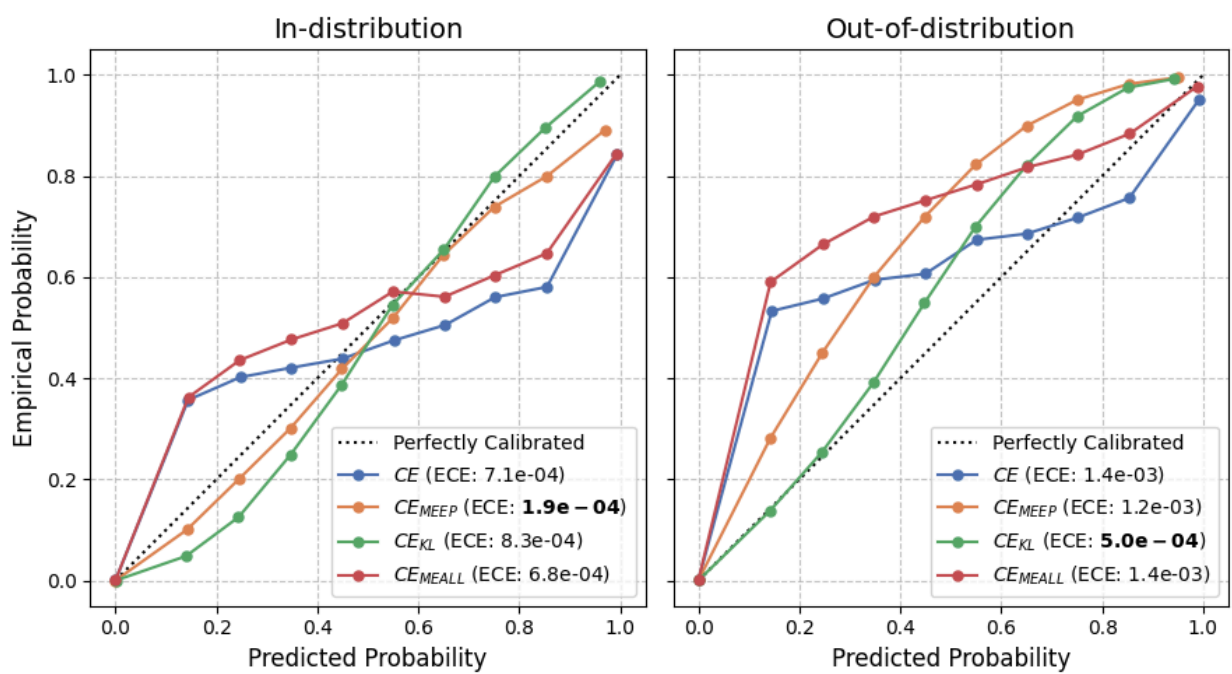

Figure 7: Reliability plots for different loss functions on ID and OOD data. Each colored line corresponds to a different loss function, with the ECE shown in parentheses (best ones are shown in bold). Points above the diagonal indicate underconfidence, while points below indicate overconfidence. A well-calibrated model should approximate the dashed diagonal line (representing perfect calibration).

# 4. Discussion

In this study, we investigated the impact of domain shift on model calibration and uncertainty estimation in white matter hyperintensity (WMH) segmentation. Our findings demonstrate that entropy-based uncertainty estimates could be used as a proxy for anticipating segmentation errors in unseen domains. Specifically, we observed a significant correlation between increasing segmentation errors due to domain shifts and rising entropy-based uncertainty estimates. By incorporating maximum-entropy regularization techniques, such as $CE_{MEEP}$ and $CE_{KL}$, we further strengthened this correlation and improved model calibration.

Our analysis also revealed that the choice of loss function significantly influences the uncertainty quantification quality. While standard cross-entropy and $CE_{MEALL}$ loss functions tend to produce lower entropy values, $CE_{MEEP}$ and $CE_{KL}$ yield higher uncertainty levels, particularly for OOD data. This suggests that $CE_{MEEP}$ and $CE_{KL}$ are more sensitive to domain shifts and better at capturing uncertainty in challenging scenarios. Additionally, our investigation into the relationship between lesion size and uncertainty revealed that smaller lesions tend to have higher uncertainty across all loss functions. This finding highlights the importance of considering lesion size when interpreting model predictions and emphasizes the need for further research into uncertainty estimation for small lesions. Models trained with maximum-entropy regularization achieved lower ECE values compared to standard training, further confirming the effectiveness of entropy-based regularization for maintaining reliable probabilistic outputs across distributions.

The analysis of prediction outcomes showed that uncertainty levels were higher for incorrect predictions (false positives and false negatives) in regularized models, especially under domain shift. This behavior is desirable in clinical practice, as it helps to identify unreliable segmentations and regions that may require expert review, enhancing the interpretability and safety of models. Notably, maximum-entropy regularization amplified uncertainty, particularly in smaller lesions, aligning model uncertainty with regions of greater clinical ambiguity. This could be valuable for detecting subtle or borderline lesions, which are typically harder to segment accurately.

In conclusion, our study underscores the importance of uncertainty estimation and model calibration in mitigating the challenges posed by domain shift in medical image analysis. By incorporating maximum-entropy regularization techniques and carefully considering the choice of loss function, more robust and reliable deep learning models for WMH segmentation can be developed. These strategies not only improve segmentation performance but also provide better indicators of prediction confidence, which are essential for safe clinical deployment in multi-center and heterogeneous imaging environments. Future work could extend this analysis by evaluating maximum-entropy regularization across different segmentation architectures, providing deeper support to the robustness and generalizability of these techniques.

## 5. Acknowledgements

The authors gratefully acknowledge NVIDIA Corporation with the donation of the GPUs used for this research, the support of Universidad Nacional del Litoral with the CAID program and Agencia Nacional de Promoción de la Investigación, el Desarrollo Tecnológico y la Innovación for the support with the PICT program. EF was supported by the Google Award for Inclusion Research (AIR) Program. VFM was partially supported by the Emerging Leaders in the Americas Program (ELAP) program. We also thank Calcul Quebec and Compute Canada.

## 6. Data Availability Statement

The datasets used in this study are publicly available:

1. The White Matter Hyperintensity (WMH) Segmentation Challenge dataset is available at https://wmh.isi.uu.nl/ under the Creative Commons Attribution-NonCommercial 4.0 International License (CC BY-NC 4.0).

2. The 3D MR Image Database of Multiple Sclerosis Patients with White Matter Lesion Segmentations (3D-MR-MS) is available at https://lit.fe.uni-lj.si/en/research/resources/3D-MR-MS/ under the Creative Commons Attribution-NonCommercial-NoDerivatives 4.0 International License (CC BY-NC-ND 4.0).

# Glossary

**White Matter Hyperintensity (WMH):** Areas of increased brightness that appear on specific types of magnetic resonance imaging (MRI) scans, indicating changes in the brain's white matter tissue. These areas are commonly associated with various neurological conditions, particularly multiple sclerosis.

**Domain Shift:** A phenomenon where the statistical properties of the data used to train a machine learning model differ from those encountered during deployment, often leading to decreased model performance.

**Model Calibration:** The extent to which a model's predicted probabilities align with observed outcomes. A well-calibrated model's confidence score accurately reflect the likelihood of correct predictions.

**Entropy:** A measure of uncertainty in probability distributions. In the context of binary segmentation, higher entropy values indicate greater uncertainty in the model's predictions.

**Overconfidence:** A situation where a model assigns very high probability values to its predictions, even when those predictions are incorrect.

**In-Distribution (ID):** Data that follows the same statistical distribution as the data used to train the model.

**Out-of-Distribution (OOD):** Data that differs significantly from the training data distribution, often leading to decreased model performance.

**Expected Calibration Error (ECE):** A metric that measures the difference between a model's predicted probabilities and the actual observed frequencies of correct predictions.

**Maximum Entropy Regularization:** A technique that encourages models to express uncertainty by penalizing low-entropy (highly confident) predictions.

**Dice Coefficient:** A metric that measures the spatial overlap between two segmentations, commonly used to evaluate the accuracy of medical image segmentation models.

**U-Net:** A specific type of convolutional neural network architecture commonly used for medical image segmentation tasks.

**Kullback-Leibler (KL) Divergence:** A measure of difference between two probability distributions.

**FLAIR (Fluid-Attenuated Inversion Recovery):** A specific type of MRI sequence that suppresses cerebrospinal fluid signals, making it easier to identify white matter lesions.

**Ground Truth:** The reference standard segmentation, typically created by expert human annotators, used to evaluate the performance of automated segmentation methods.

## CRediT Author Statement

**Franco Matzkin:** Conceptualization, Methodology, Software, Validation, Formal Analysis, Investigation, Data Curation, Writing - Original Draft, Writing - Review & Editing, Visualization

**Diego H. Milone:** Supervision, Conceptualization, Methodology, Writing - Review & Editing, Project Administration

**José Dolz:** Resources, Supervision, Writing - Review & Editing, Methodology

**Agostina Larrazabal:** Methodology, Software

**Enzo Ferrante:** Supervision, Conceptualization, Methodology, Writing - Review & Editing, Project Administration

# Funding

This work was supported by the National Scientific and Technical Research Council (CONICET, Argentina) and the Emerging Leaders in the Americas Program (ELAP) from the Government of Canada, which funded a research stay at ETS Montreal. The funding sources had no involvement in the study design; in the collection, analysis and interpretation of data; in the writing of the report; or in the decision to submit the article for publication.

# Declaration of Generative AI Use in Scientific Writing

The authors declare the use of large language models (ChatGPT and Claude.ai) solely for grammar checking and language translation assistance, as Spanish is the native language of the research team. All scientific content, analysis, and conclusions were independently developed by the authors. The final manuscript was thoroughly reviewed and approved by all authors to ensure accuracy and integrity of the scientific content.

# Ethics Statement

No ethical approval was required for this study as it utilized only publicly available datasets: the White Matter Hyperintensity (WMH) Segmentation Challenge dataset and the 3D MR Image Database of Multiple Sclerosis Patients (3D-MR-MS), both of which are freely accessible for research purposes under their respective Creative Commons licenses.